\documentclass[seceq]{ptptex}

\usepackage{graphicx}
\usepackage{multirow}

\newcommand{\ket}[1]{\left|{#1}\right\rangle}

\newcommand{\ph}{\varphi}

\notypesetlogo                       

\title{A short introduction to Fibonacci anyon models}

\author{Simon Trebst$^{(1)}$, Matthias Troyer$^{(2)}$, Zhenghan Wang$^{(1)}$, Andreas W. W. Ludwig$^{(3)}$}
\inst{$^{(1)}$Microsoft Research, Station Q, \\
University of California, Santa Barbara, CA 93106 \\ 
$^{(2)}$ Theoretische Physik, ETH Zurich, 8093 Zurich, Switzerland \\
$^{(3)}$ Physics Department, University of California, Santa Barbara, California 93106
}

\recdate{May, 2008}

\abst{
We discuss how to construct models of interacting anyons by generalizing quantum spin
Hamiltonians to anyonic degrees of freedom.
The simplest interactions energetically favor pairs of anyons to fuse into the trivial (``identity") 
channel, similar to the quantum Heisenberg model favoring pairs of spins to form spin singlets. 
We present an introduction to the theory of anyons and discuss in detail how basis sets and 
matrix representations of the interaction terms can be obtained, using non-Abelian Fibonacci 
anyons as example. 
Besides discussing the ``golden chain'', a one-dimensional system 
of anyons with nearest neighbor interactions, we also present
the derivation of more complicated interaction terms, such as three-anyon interactions in the spirit 
of the Majumdar-Ghosh spin chain, longer range interactions and two-leg ladders. 
We also discuss generalizations to anyons with general non-Abelian SU(2)$_k$ statistics. 
The $k\to \infty$ limit of the latter yields ordinary SU(2) spin chains.
}

\begin{document}

\maketitle

\section{Introduction}

While in classical mechanics the exchange of two identical particles
does not change the underlying state, quantum mechanics allows for 
more complex behavior. In three-dimensional quantum systems the 
exchange of two identical particles may result in a sign-change of the
wavefunction which 
distinguishes
fermions from bosons. 
Two-dimensional quantum systems -- such as electrons confined 
between layers of semiconductors -- can give rise to exotic particle
statistics, where the exchange of two identical (quasi)particles can
in general be described by either Abelian or non-Abelian statistics.
In the former, the exchange of two particles gives rise to a complex
phase $e^{i\theta}$, where $\theta=0,\pi$ correspond to the statistics 
of bosons and fermions, respectively, and $\theta \neq 0,\pi$ is referred
to as the statistics of Abelian {\em anyons}.
The statistics of non-Abelian anyons are described by $k\times k$ 
unitary matrices acting on a degenerate ground-state manifold with 
$k>1$. In general, two such unitary matrices $A, B$ do not necessarily 
commute, i.e. $AB\neq BA$, or in
more 
mathematical language,
the $k\times k$ 
unitary matrices form a non-Abelian group when $k>1$, hence the 
term non-Abelian anyons.

Anyons appear as emergent quasiparticles in fractional quantum 
Hall states and as excitations in microscopic models of 
frustrated quantum magnets that harbor topological quantum liquids
\cite{Kitaev03,Kitaev06,LevinWen05}.
While for most quantum Hall states the exchange statistics is Abelian,
there are quantum Hall states at certain filling fractions, 
e.g.  $\nu=\frac{5}{2}$ and $\nu=\frac{12}{5}$, 
for which non-Abelian quasiparticle statistics have been proposed
\cite{MooreRead,ReadRezayi},
namely those of so-called Ising anyons \cite{Nayak96} and 
Fibonacci anyons \cite{Slingerland01} respectively.
Non-Abelian anyons have also generated considerable interest in
proposals for topological quantum computation \cite{Nayak08}, where braiding
of anyons is used to perform the unitary transformations of a 
quantum computation. 
The simplest anyons with non-Abelian braiding statistics that can 
give rise to universal quantum computation\footnote{
Roughly speaking, a universal quantum computer is a general-purpose 
quantum computer which is capable of simulating any program on 
another quantum computer.}
are the so-called Fibonacci anyons which we will discuss in detail in this 
manuscript.

In the following, we will first give a short 
introduction to
the mathematical
theory of anyons, and discuss how to (consistently) describe the degenerate 
manifold of a set of (non-interacting) anyons. Having established the basic formalism
we will then turn to the question of how to model interactions between anyons
and explicitly construct matrix representations of generalized quantum spin
Hamiltonians. 
We then discuss an alternative formulation in terms of non-Abelian SU(2)$_k$  anyons.
Rounding off the manuscript, we shortly review some recent work analyzing 
the ground-state phase diagrams of these Hamiltonians.

\section{Basic theory}

\subsection{Algebraic theory of anyons}

In general terms, we can describe anyons by a mathematical 
framework called tensor category theory.
In such a categorical description, anyons are simple objects in
the corresponding tensor categories, and anyon types are the
isomorphism classes of anyons.
Here we will not delve into this difficult mathematical subject, but
focus on the theory of Fibonacci anyons, where many simplifications 
occur.

\subsection{Particle types and fusion rules}
To describe a system of anyons, we list the species of the anyons in the
system, also called the particle types or topological charges or simply
labels (and many other names); we also specify the anti-particle type of
each particle type.  We will list the particle types as
$\{x_i\}_{i=0}^{n-1}$, and use $\{X_i\}_{i=0}^{n-1}$ to denote a
representative set of anyons, where the type of $X_i$ is $x_i$.

In any anyonic system, we always have a trivial particle type
denoted by $\bf{1}$, which represents the ground states of the 
system or the vacuum. In the list of particle types above, we assume
$x_0=\bf{1}$. The trivial particle is its own anti-particle. The
anti-particle of $X_i$, denoted as $X_i^{*}$, is always of the
type of another $X_j$.  If $X_i$ and $X_i^{*}$ are of the same
type, we say $X_i$ is self-dual.

To have a non-trivial anyonic system, we need at least one more
particle type besides $\bf{1}$. The Fibonacci anyonic system is such an
anyonic system with only two particle types: the trivial type
$\bf{1}$, and the nontrivial type $\tau$. Anyons of type $\tau$
are called the Fibonacci anyons. Fibonacci anyons are self-dual:
the anti-particle type of $\tau$ is also $\tau$. Strictly
speaking, we need to distinguish between anyons and their types.
For Fibonacci anyons, this distinction is unnecessary.  Therefore,
we will refer to $\tau$ both as an anyon and its type, and no
confusion should arise.

Anyons can be combined in a process called the fusion of anyons,
which is similar to combining two quantum spins to form a new total spin. 
Repeated fusions of the same two anyons do not necessarily 
result in an anyon of the same type: 
the resulting anyons may be of several different types
each with certain probabilities (determined by the theory). In
this sense we can also think of fusion as a measurement. 
It follows that given two
anyons $X,Y$ of type $x,y$, the particle type of the fusion,
denoted as $X\otimes Y$, is in general not well-defined.

Given an anyon $X$, if the fusion of $X$ with any other anyon $Y$
(maybe $X$ itself) always produces an anyon of the same type, then
$X$ is called an Abelian anyon.  If neither $X$ nor $Y$ is
Abelian, then there will be anyons of more than one type as the
possible fusion results. When such fusion occurs, we say that the
fusion has {\em multi-fusion channels}.

Given two anyons $X,Y$, we formally write the fusion result as
$X\otimes Y\cong \oplus_i n_iX_i$, where $X_i$ are all anyons in a
representative set, and $n_i$ are non-negative integers.  The
non-negative integer $n_i$ is called the multiplicity of the
occurrence of anyon $X_i$. Multi-fusion channels correspond to
$\sum_i n_i>1$. Given an anyonic system with anyon representative
set $\{X_i\}_{i=0}^{n-1}$, then we have $X_i\otimes X_j\cong
\oplus_{k=0}^{n-1} N_{i,j}^k X_k$, or equivalently, $x_i\otimes
x_j=\oplus_{k=0}^{n-1}N_{i,j}^k x_k$. The non-negative integers
$N_{i,j}^k$ are called the fusion rules of the anyonic system.
If $N_{i,j}^k \neq 0$, we say the fusion of $X_i$ and $X_j$ to
$X_k$ is admissible.

The trivial particle is Abelian as the fusion of the trivial
particle with any other particle $X$ does not change the type of
$X$, i.e., ${\bf{1}} \otimes x =x$ for any type $x$.

For the Fibonacci anyonic system the particle types are denoted
as $\bf{1}$ and $\tau$, and the fusion rules are given by: 
\begin{eqnarray}
\bf{1}\otimes \tau & = & \tau \nonumber \\ 
\tau\otimes \bf{1} & = & \tau \nonumber \\ 
\tau\otimes \tau & = & \bf{1} \oplus \tau \nonumber \,, 
\end{eqnarray}
where the $\oplus$ denotes the two possible fusion channels.

\subsection{Many anyon states and fusion tree basis}

\begin{figure}[h]
  \center{
      \includegraphics[width=0.75\columnwidth]{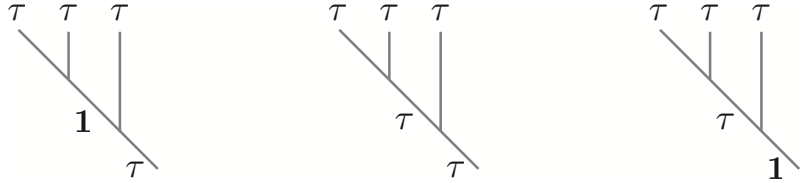}
  }
  \caption{Fusion trees of three Fibonacci anyons (top row).}
  \label{Fig:FusionTree}
\end{figure}

A defining feature of non-Abelian anyons is the existence of
multi-fusion channels. Suppose we have three $\tau$ anyons
localized in the plane, well-separated, and numbered as $1,2,3$.
We would like to know when all three anyons are brought together to fuse,
what kinds of anyons will this fusion result in? When anyons $1$ and 
$2$ are combined, we may see $\bf{1}$ or $\tau$.  If the resulting anyon
were $\bf{1}$, then after combining with the third $\tau$, we
would have a $\tau$ anyon.  If the resulting anyon were $\tau$,
then fusion with the third anyon would result in either $\bf{1}$
or $\tau$. Hence the fusion result is not unique. Moreover, even if
we fix the resulting outcome as $\tau$, there are still two
possible fusion paths: the first two $\tau$'s were fused to
$\bf{1}$, then fused with the third $\tau$ to $\tau$, or the first two
$\tau$'s were fused to $\tau$, then fused with the third $\tau$ to $\tau$. 
Each such fusion path will be recorded by a graphical notation
of the {\em fusion tree}, see Fig.~\ref{Fig:FusionTree}.

A {\em fusion path} is a labeling of the fusion tree where each edge is
labeled by a particle type, and the three labels around any
trivalent vertex represent a fusion admissible by the fusion
rules.  If not all particles are self-dual, then the edges of the
fusion tree should be oriented.  We always draw anyons to be fused
on a straight line, and the fusion tree goes downward. The top
edges are labeled by the anyons to be fused, and the bottom edge
represents the fusion result and is also called the total charge
of the fused anyons.

In general, given $n$ $\tau$-anyons in the plane localized at
certain well separated places, we will assume the total charge at
the $\infty$ boundary is either $\bf{1}$ or $\tau$. In theory any
superposition of $\bf{1}$ or $\tau$ is possible for the total
charge, but it is physically reasonable to assume that such
superpositions will decohere into a particular anyon if left
alone. Let us arrange the $n$ anyons on the real axis of the
plane, numbered as $1,2,\cdots, n$.  When we fuse the anyons
$1,2,\cdots, n$ consecutively, we have a fusion tree as below:

\begin{figure}[h]
  \center{
      \includegraphics[width=0.4\columnwidth]{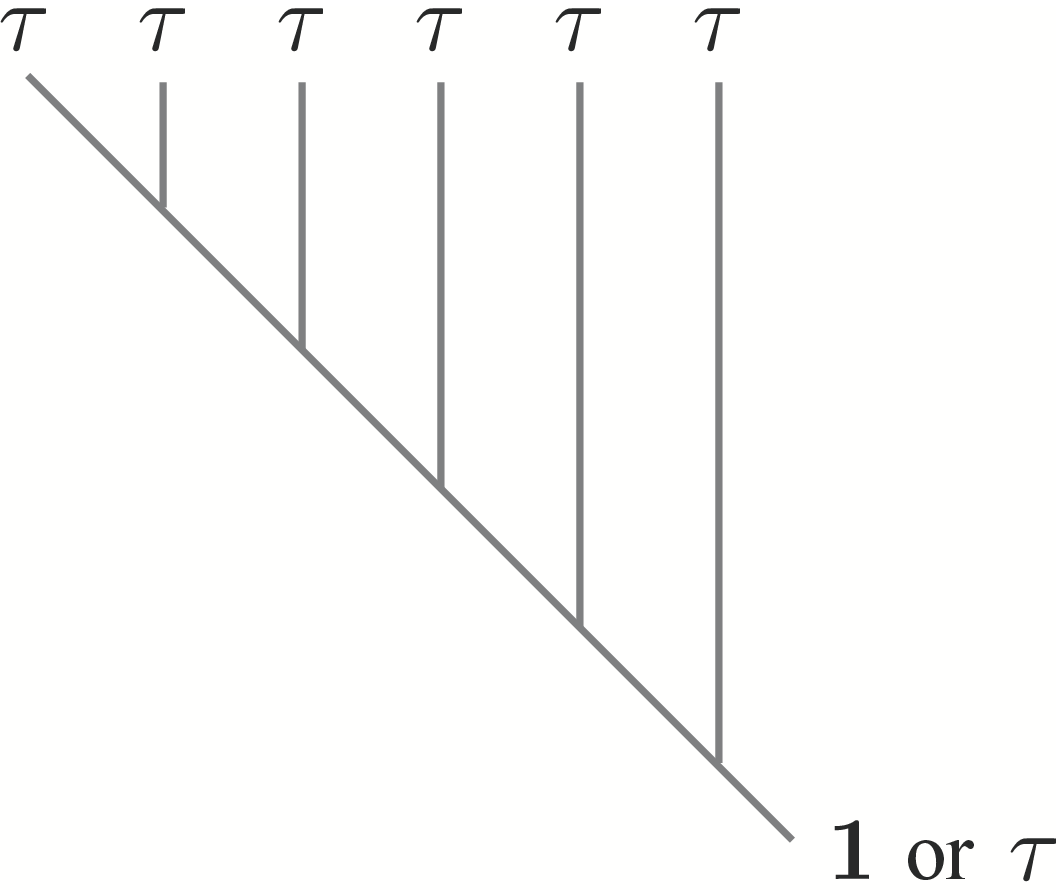}
  }
\end{figure}

The ground-state manifold of a multi-anyon system in the plane
even when the positions of the anyons are fixed might be
degenerate: 
there is more than one ground state
(in reality the
energy differences between the different ground states go to $0$ 
exponentially as the anyon separations go to infinity; we will ignore 
such considerations here, and always assume that anyons are well 
separated until they are brought together for fusion.) 
Such a degeneracy is in fact necessary for non-Abelian statistics to 
occur. How can we describe a basis for this degenerate ground state
manifold?

As we see in the example of three $\tau$ anyons, there are
multi-fusion paths, which are represented by labelings of the
fusion tree. We claim that these fusion paths represent an
orthonormal basis of the degenerate ground-state manifold.\footnote{
We will not further justify this assertion, but mention that in the
conformal field theory (CFT) description of fractional quantum Hall 
liquids the ground states can be described by conformal blocks,
which form a basis of the modular functor. 
Conformal blocks are
known to be represented by labeled fusion trees, which we refer 
to as fusion paths.}

The fusion tree basis of a multi-anyon system then leads to a
combinatorial way to compute the degeneracy: count the number of
labelings of the fusion tree or equivalently the number of fusion paths.  
Consider $n$ $\tau$-anyons in the
plane with total charge $\tau$, and denote the ground state
degeneracy as $F_n$. Simple counting shows that $F_0=0$ and
$F_1=1$. Easy induction then gives $F_{n+1}=F_n+F_{n-1}$.  This is
exactly the Fibonacci sequence, hence the name of Fibonacci
anyons.

As alluded to above,
when two $\tau$ anyons are fused, $\bf{1}$ and
$\tau$ each occurs with a certain probability. This probability is
given by the so-called {\em quantum dimension} of an anyon.  Consider the
fusion coefficients $N_{i,j}^k$ of a theory, if we regard the
particle types $x_i$ as variables and the fusion rules as
equations for $x_i$.  Then in a unitary theory the solutions $d_i$ of $x_i$
which are $\geq 1$ are the quantum dimensions of the anyons of type $x_i$.
$d_i$ is also the Perron-Frobenius eigenvalue of the matrix $N_i$
whose $(j,k)$th entry is $N_{i,j}^k$.  We also introduce the total
quantum order $D=\sqrt{\sum_i d_i^2}$. The quantum dimension of
the trivial type $\bf{1}$ is always $d_0=1$. In the Fibonacci
theory, the quantum dimension of $\tau$ is the golden ratio
$\ph=\frac{1+\sqrt{5}}{2}$.  When two $\tau$ anyons fuse, the
probability to see $\bf{1}$ is $p_0=\frac{1}{\ph^2}$, and the
probability to see $\tau$ is $p_1=\frac{\ph}{\ph^2}=\frac{1}{\ph}$.

\subsection{F-matrices and pentagons}

\begin{figure}[h]
  \center{
      \includegraphics[width=0.5\columnwidth]{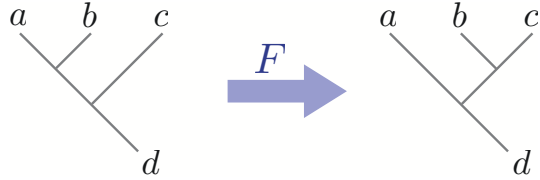}
  }
  \caption{(color online)
  		The two fusion trees of three anyons that both result in the same anyon $d$
                  are related by an ``$F$-move".}
  \label{Fig:F-move}
\end{figure}

In the discussion of the fusion tree basis above, we fuse the anyons
$1,2,\cdots, n$ consecutively one by one from left to right, e.g., $n=3$
gives the left fusion tree below.  
We may as well choose any other order for the fusions.  
For example, in the case of
three $\tau$'s with total charge $\tau$, we may first fuse the
second and third $\tau$'s, then fuse the resulting anyon with the first
$\tau$. This will lead to the fusion tree on the right as shown 
in Fig.~\ref{Fig:F-move}.

Given $n$ anyons with a certain total charge, then each order of
the fusions is represented by a fusion tree, and the admissible
labelings of the respective fusion trees each constitute a basis of 
the multi-anyon system.

The change from the left fusion tree to the right fusion tree 
in Fig.~\ref{Fig:F-move} is called the
$F$-move.  Since both fusion tree bases describe the same
degenerate ground state manifold of $3$ anyons with a certain
total charge, they should be related by a unitary transformation.
The associated unitary matrix is called the $F$-matrix.  
The $F$-matrix will be denoted as $F^{abc}_d$, where $a,b,c$ are the
anyons to be fused, and $d$ is the resulting anyon or total charge
(Complications from fusion coefficients $N_{i,j}^k >1$ are ignored.)

\begin{figure}[h]
  \center{
      \includegraphics[width=0.75\columnwidth]{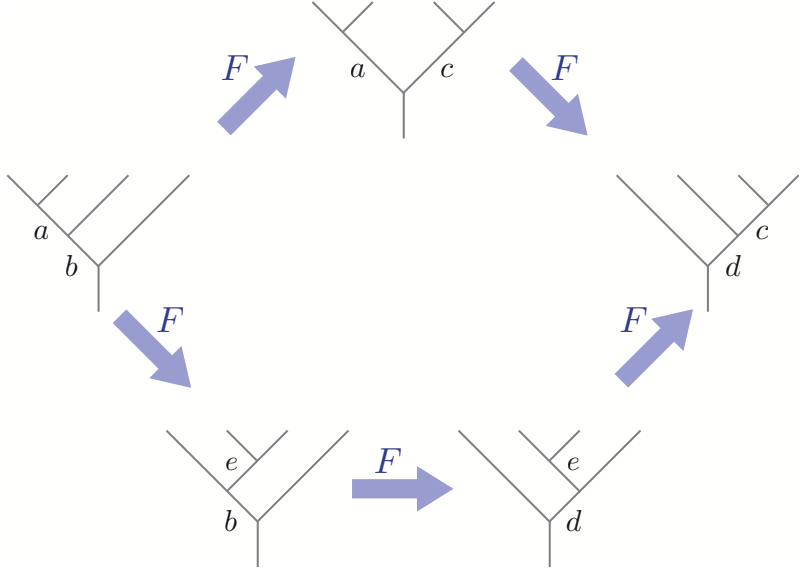}
  }
  \caption{(color online) The pentagon relation for the ``$F$-moves".}
  \label{Fig:pentagon}
\end{figure}

For more than three anyons, there will be many more different fusion trees.
To have a consistent theory, a priori we need to specify the
change of basis matrices for any number of anyons in a consistent
way: for example as shown in Fig.~\ref{Fig:pentagon} the left-most and 
right-most fusion trees of four anyons can be related to each other by 
$F$-moves in two different sequences of applications of $F$-moves.

Fortunately, a mathematical theorem guarantees that the consistency
equations for the above fusion trees, called the pentagons, are
all the equations that need to be satisfied, i.e., all other
consistencies are consequences of the pentagons.
Note that the pentagons are just polynomial equations for the 
entries of the $F$-matrices.

To set up the pentagons, we need to explain the consistency of
fusion tree bases for any number of anyons.  Consider a fusion tree $T$,
and a decomposition of $T$ into two sub-fusion trees $T_1, T_2$ by
cutting an edge; the resulting new edge of $T_1, T_2$ will also be
referred to as edge $e$. The fusion tree basis for $T$ has a
corresponding  decomposition: if $x_i$'s are the particle types of
the theory (we assume they are all self-dual), for each $x_i$, we
have a fusion tree basis for $T_1,T_2$ with the edge $e$ labeled by
$x_i$.  Then the fusion tree basis of $T$ is the direct sum over all $x_i$
of the tensor product: (the fusion tree basis of $T_1$) $\otimes$ (the fusion tree
basis of $T_2$).

In the pentagons, an $F$-move is applied to part of the fusion
trees in each step. The fusion tree decomposes into two pieces:
the part where the $F$-move applies, and the remaining part.  It
follows that the fusion tree basis decomposes as a direct sum of
two terms: corresponding to $\bf{1}$ and $\tau$.

Given a set of fusion rules $N_{i,j}^k$ solving the pentagons
turns out to be a difficult task (even with the help of computers).  
However, certain normalizations can be made to simplify the solutions. 
If one of the indices of 
the $F$-matrix
$a,b,c$ is the trivial type $\bf{1}$, 
we may assume $F^{a,b,c}_d=1$.  
In the Fibonacci theory, we may also assume $F^{a,b,c}_{\bf{1}}=1$. It
follows that the only non-trivial $F$-matrix is
$F^{\tau,\tau,\tau}_{\tau}$, which is a $2\times 2$ unitary
matrix.

There are many pentagons even for the Fibonacci theory depending
on the four anyons to be fused and their total charges: a priori
$2^5 = 32$.  It is easy to see that the only non-trivial pentagon for
$F$ is the one with $5$ $\tau$'s at all outer edges.  The pentagon
is a matrix equation for $F$ extended to a bigger Hilbert space.
To write down the pentagon, we need to order the fusion tree basis
with respect to the decomposition above carefully.

Written explicitly for Fibonacci anyons the pentagon equation reads
\begin{equation}
\left (F^{\tau\tau c}_\tau \right)_a^d \left (F^{a\tau\tau}_\tau \right)_b^c = \left (F^{\tau\tau\tau}_d \right)_e^c\left (F^{\tau e\tau}_\tau \right)_b^d\left (F^{\tau\tau\tau}_b \right)_a^e \,,
\end{equation}
where the indices $a,b,c,d,e$ label the inner edges of the fusion tree as shown
in Fig.~\ref{Fig:pentagon}.
There are only a few different matrices appearing, of which four are uniquely determined by the fusion rules 
\begin{equation}
F^{\tau\tau\tau}_1=F^{1\tau\tau}_\tau=F^{\tau1\tau}_\tau=F^{\tau\tau1}_\tau=1
\end{equation}
in a basis $\{1,\tau\}$ for the labeling on the central edge. The only nontrivial matrix is $F^{\tau\tau\tau}_\tau$. Setting $b=c=1$ the pentagon equation simplifies to
\begin{equation}
\left (F^{\tau\tau\tau}_\tau \right)_1^1 = \left (F^{\tau\tau\tau}_\tau \right)_\tau^1\left (F^{\tau\tau\tau}_\tau \right)_1^\tau \,,
\end{equation}
which combined with the condition that $F^{\tau\tau\tau}_\tau$ is unitary constrains the matrix, up to arbitrary phases, to be
\begin{equation}
F^{\tau\tau\tau}_\tau={F^{\tau\tau\tau}_\tau}^{\dag}  =  \left(\begin{array}{cc}\ph^{-1}& \ph^{-1/2}\\ \ph^{-1/2}&-\ph^{-1}\end{array} \right) \,,
\label{eq:F}
\end{equation}
where $\ph=(\sqrt{5}+1)/2$ is the golden ratio.

\subsection{R-matrix and hexagons}

Given $n$ anyons $Y_i$ in a surface $S$, well-separated at fixed
locations $p_i$, we may consider the ground states $V(S;p_i,Y_i)$
of this quantum system.  Since an energy gap in an anyonic system
is always assumed, if two well-separated anyons $Y_i, Y_j$ are
exchanged slowly enough, the system will remain in the ground states
manifold $V(S;p_i,Y_i)$. If $\ket{\Psi_0} \in V(S;p_i,Y_i)$ is the initial
ground state, then after the exchange, or the braiding of the two
anyons $Y_i, Y_j$ in mathematical parlor, the system will be in
another ground state $\ket{\Psi_1}=\sum_i b_i e_i$ in $V(S;p_i,Y_i)$,
where $e_i$ is an othonormal basis of the ground states
manifold $V(S;p_i,Y_i)$. When $\ket{\Psi_0}$ runs over the basis $e_i$,
we obtain a unitary matrix $R_{i,j}$ from $V(S;p_i,Y_i)$ to
itself.  In mathematical terms, we obtain a representation of the
mapping class group of the punctured surface $S$. If $S$ is the disk, the
mapping class group is called the braid group.  In a nice basis of
$V(S;p_i,Y_i)$, the braiding matrix $R_{i,j}$ becomes diagonal.

To describe braidings carefully, we introduce some conventions.
When we exchange two anyons $a,b$ in the plane, there are two
different exchanges which are not topologically equivalent: their
world lines are given by the following two pictures, which are
called braids mathematically. In our convention
time goes
upwards.  When we exchange two anyons, we will refer to the right
process, which is called the right-handed braiding. The left process
is the inverse, left-handed braiding.

\begin{figure}[h]
  \center{
      \includegraphics[width=0.18\columnwidth]{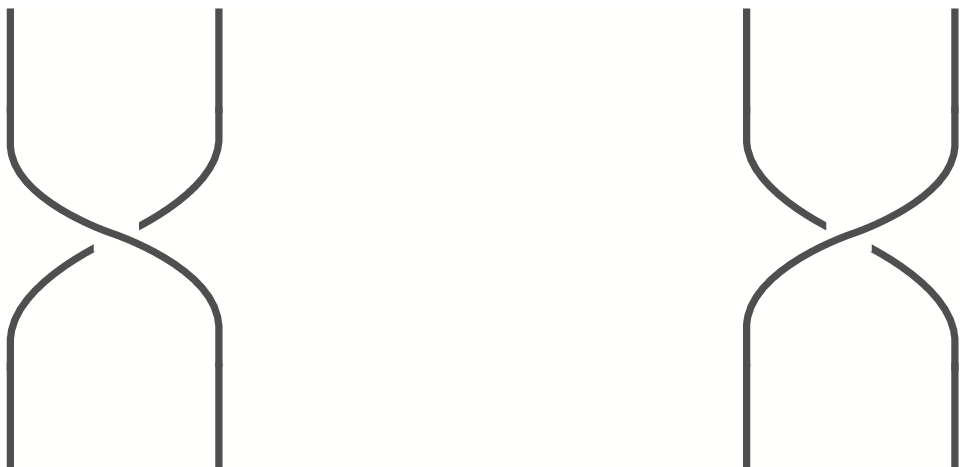}
  }
\end{figure}

Now a comment about fusion trees is necessary.  
In our convention, we draw the fusion trees
downwards.  If we want to interpret a fusion tree as a physical
process in time, we should also introduce the conjugate operator of the
fusion: splitting of anyons from one to two. Then as time goes
upwards, a fusion tree can be interpreted as a splitting of one anyon 
into many.

All the braiding matrices can be obtained from the $R$-matrices
combined with $F$-matrices. Let $V^{a,b}_c$ be the ground state
manifold of two anyons of types $a,b$ with total charge $c$. Let
us assume all spaces $V^{a,b}_c$ are one-dimensional, and
$e^{a,b}_c$ be its fusion tree basis.

When anyons $a$ and $b$ are braided by $R_{a,b}$, the state $e^{a,b}_c$ in
$V^{a,b}_c$ is changed into a state $R_{a,b}e^{a,b}_c$ in
$V^{b,a}_c$. Since both $R_{a,b}e^{a,b}_c$ and $e^{b,a}_c$ are
non-zero vectors in a one-dimensional Hilbert space $V^{b,a}_c$,
they are equal up to a phase, denoted as $R^{b,a}_c$, i.e,
$R_{a,b}e^{a,b}_c=R^{b,a}_c e^{b,a}_c$.  Here, $R^{b,a}_c$ is a
phase, but in general, $R^{b,a}_c$ is a unitary matrix. We should
mention that in general $R^{b,a}_c$ is not the inverse of
$R^{a,b}_c$.  Their product involves the twists of particles.

\begin{figure}[h]
  \center{
      \includegraphics[width=0.5\columnwidth]{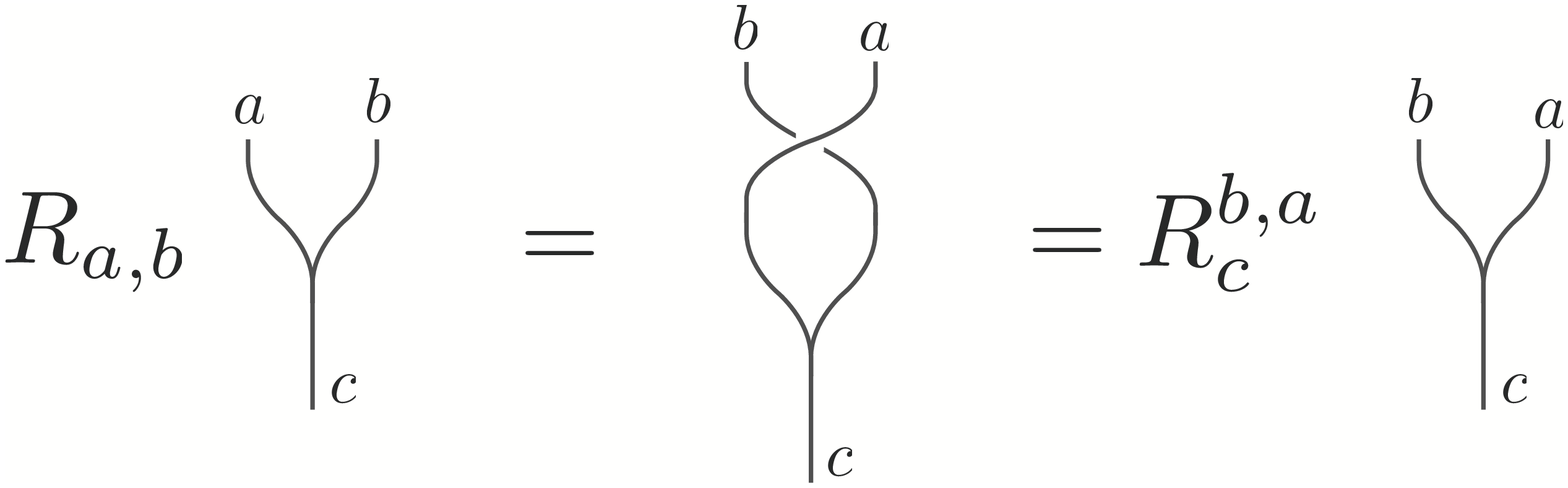}
  }
\end{figure}

As we have seen before anyons can be fused or splitted, therefore
braidings should be compatible with them.  For example, given two
anyons $c,d$, we may first split $d$ to $a,b$, then braid $c$ with
$a$ followed by braid $c$ with $b$, or we may braid $c$ and $d$
first, then split $d$ into $a,b$. These two processes are
physically equivalent, therefore their resulting matrices should
be the same. 
%
%
Applying the two operators on the fusion tree basis $e^{c,d}_m$,
we have an identity in pictures:
\begin{figure}[h]
  \center{
      \includegraphics[width=0.35\columnwidth]{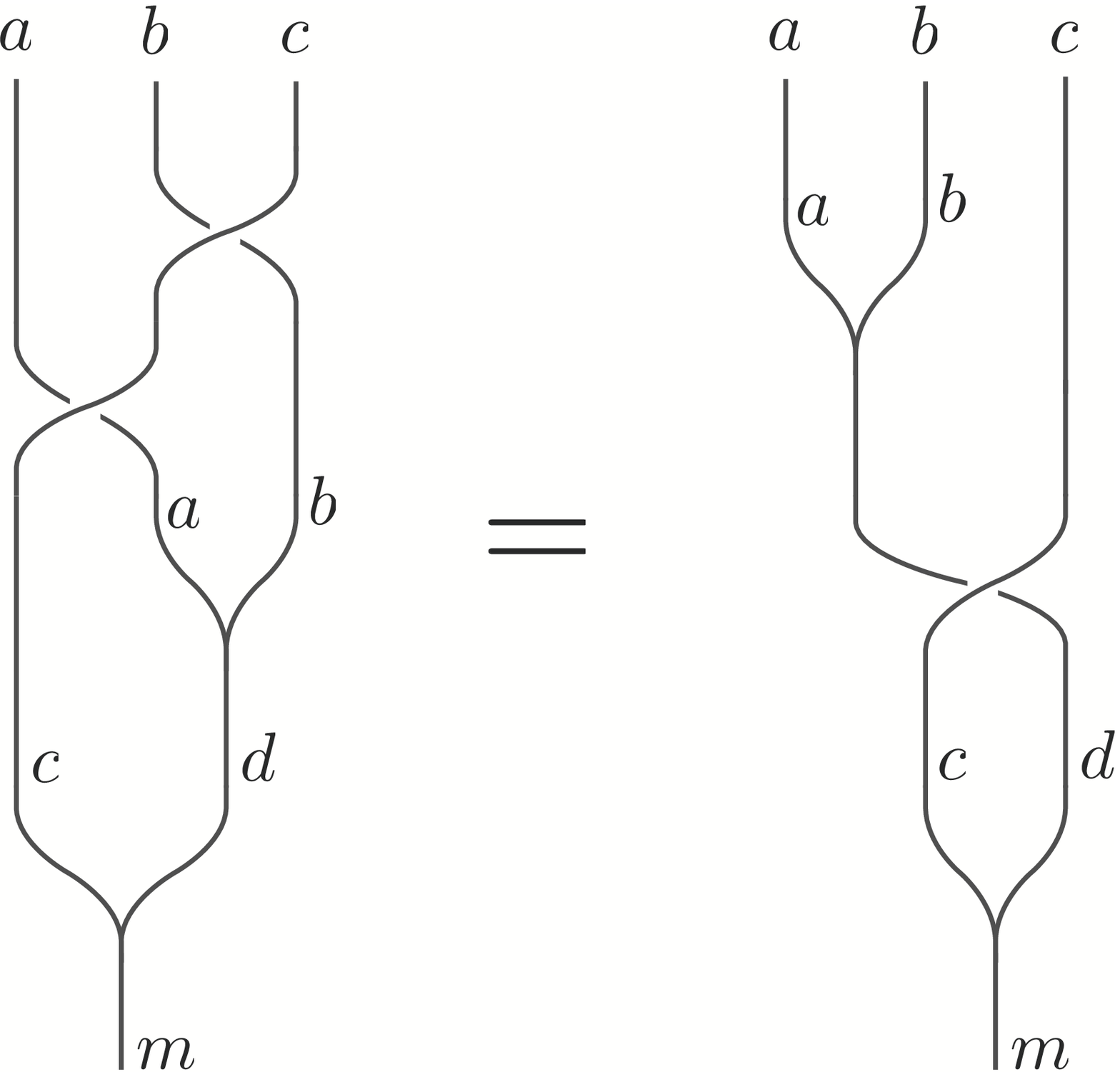}
  }
\end{figure}

\begin{figure}[t]
  \center{
      \includegraphics[width=0.9\columnwidth]{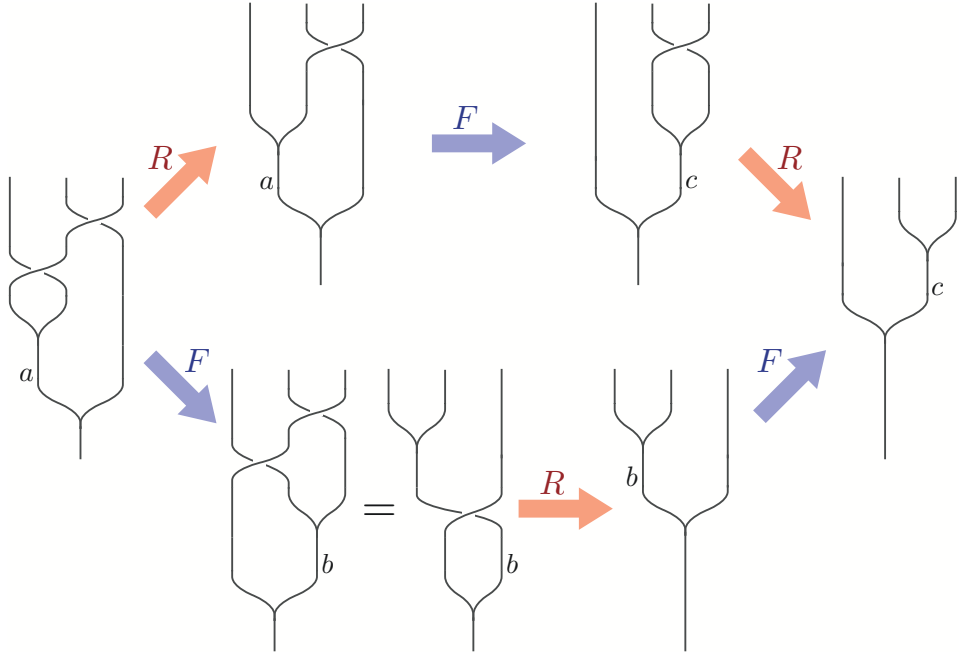}
  }
  \caption{(color online) The hexagon relation for ``$R$-moves" and ``$F$-moves".}
  \label{Fig:hexagon}
\end{figure}
The same identity can be also obtained as a composition of
$F$-moves and braidings as shown in Fig.~\ref{Fig:hexagon}.
It follows the composition of the $6$ matrices, hence the name
hexagon, should be the same as the identity.  The resulting
equations are called hexagons.  There is another family of
hexagons obtained by replacing all right-handed braids with
left-handed ones.  In general, these two families of hexagons are
independent of each other. 
Similar to the pentagons, a mathematical theorem says that the
hexagons imply all other consistency equations for braidings.

Written explicitly for Fibonacci anyons the hexagon equation reads
\begin{equation}
R^{\tau,\tau}_c\left( F^{\tau\tau\tau}_\tau\right) _a^c R^{\tau,\tau}_a= \sum_b \left( F^{\tau\tau\tau}_\tau\right)_b^c R^{\tau,b}_\tau \left( F^{\tau\tau\tau}_\tau\right) _a^b \,,
\end{equation}
where again the indices $a,b,c$ label the internal edges of the fusion
trees as shown in Fig.~\ref{Fig:hexagon}.
Inserting the $F$-matrix (\ref{eq:F}) and realizing that braiding a particle 
around the trivial one is trivial: $R^{\tau,1}_\tau = R^{1,\tau}_\tau = 1$ the 
hexagon equation becomes
\begin{equation}
\left(
\begin{array}{ll}
(R^{\tau,\tau}_1)^2\ph^{-1} & R^{\tau,\tau}_1 R^{\tau,\tau}_\tau \ph^{-1/2} \\
R^{\tau,\tau}_1 R^{\tau,\tau}_\tau \ph^{-1/2}  & -(R^{\tau,\tau}_\tau)^2\ph^{-1}
\end{array}
\right)=
\left(
\begin{array}{ll}
R^{\tau,\tau}_\tau\ph^{-1}+ \ph^{-2} &  (1-R^{\tau,\tau}_\tau)\ph
   ^{-3/2} \\
(1-R^{\tau,\tau}_\tau) \ph ^{-3/2} &
 R^{\tau,\tau}_\tau\ph^{-2}+  \ph^{-1}
\end{array}
\right) \,,
\end{equation}
which has the solution
 \begin{equation}
 R^{\tau\tau}_1= e^{+4\pi i/5},\qquad R^{\tau\tau}_\tau= e^{-3\pi i/5}.
 \end{equation}
%
%
 The combined operation of a basis transformation $F$ before applying the $R$-matrix 
 is often denoted by the braid-matrix $B$
 \begin{equation}
 B=F^{a\tau\tau}_cR_{\tau,\tau}F^{a\tau\tau}_c.
 \end{equation}
 Using a basis $\{|abc\rangle\}$ for the labelings adjacent to the two anyons to be braided the basis before the basis transformation is 
 \begin{equation}
 \{|1\tau1\rangle, |\tau\tau1\rangle, |1\tau\tau\rangle,|\tau1\tau\rangle,|\tau\tau\tau\rangle\}
 \label{eq:basisbefore}
\end{equation}
and after the basis change to a basis $\{|a\tilde{b}c\rangle\}$ using an $F$ matrix the basis is
 \begin{equation}
 \{|111\rangle, |\tau\tau1\rangle, |1\tau\tau\rangle,|\tau1\tau\rangle,|\tau\tau\tau\rangle\}.
 \label{eq:basisafter}
 \end{equation}
 as illustrated here.
\begin{figure}[h]
  \center{
      \includegraphics[width=0.75\columnwidth]{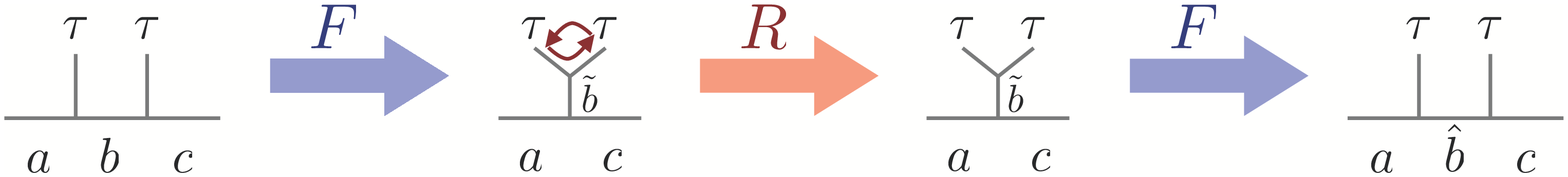}
  }
\end{figure}

In this representation the $F$-matrix is given by
\begin{equation}
F = \left(\begin{array}{ccccc}
1&&&& \\
&1&&& \\
&&1&&\\
&&&\ph^{-1}&\ph^{-1/2}\\
&&&\ph^{-1/2}&-\ph^{-1}\end{array}
 \right)
\end{equation}
and the $R$-matrix is 
\begin{equation}
R={\rm diag}(e^{4\pi i/5},e^{-3\pi i/5},e^{-3\pi i/5},e^{4\pi i/5},e^{-3\pi i/5}).
\end{equation}
We finally obtain for the braid matrix
\begin{equation}
B=FRF^{-1} = 
 \left(\begin{array}{ccccc}
e^{4\pi i/5}&&&& \\
&e^{-3\pi i/5}&&& \\
&&e^{-3\pi i/5}&&\\
&&&\ph^{-1}e^{-4\pi i/5}&-\ph^{-1/2}e^{-2\pi i/5}\\
&&&-\ph^{-1/2}e^{-2\pi i/5}&-\ph^{-1}\end{array}
 \right) \,.
\end{equation}

With the explicit matrix representations of the basis transformation $F$ and the
braid matrix $B$, we are now fully equipped to derive matrix representations of 
Hamiltonians describing interactions between Fibonacci anyons. 

\section{Hamiltonians}

Considering a set of Fibonacci anyons we will now address how to model interactions
between these anyons. Without any interactions, the collective state of the anyons will simply 
be described by the large degenerate manifold described by the Hilbert space introduced above. 
If, however, the anyons interact, this degeneracy will be split and a non-trivial collective 
ground state is formed.
In this section we will first motivate a particular type of interaction, which generalizes 
the well-known Heisenberg exchange interaction to anyonic degrees of freedom, and
then explicitly derive various Hamiltonians of interacting Fibonacci anyons that 
correspond to well-known models of SU(2) spin-1/2's.

Two SU(2) spin-1/2's can be combined to form a total spin singlet $\bf 0$ or a total
spin triplet $\bf 1$, which in analogy to the anyonic fusion rules we might write as
\begin{equation}
1/2 \otimes 1/2 = {\bf 0} \oplus {\bf 1}\,. \nonumber
\end{equation}
If the two spins are far apart and interact only weakly, these two states are degenerate.
However, if we bring the two spins close together a strong exchange will be mediated
by a virtual tunneling process and the
degeneracy between the
 two total spin states will be lifted.
This physics is captured by the Heisenberg Hamiltonian which for SU(2) spins is given by
\begin{equation}
  \mathcal{H}^{\rm SU(2)}_{\rm Heisenberg} 
  = J \sum_{\langle ij \rangle} \vec{S}_i \cdot \vec{S}_j 
  = \frac{J}{2} \sum_{\langle ij \rangle}  \left( \vec{T}^2_{ij} - \vec{S}^2_i - \vec{S}^2_j \right)
  = \frac{J}{2} \left( \sum_{\langle ij \rangle} \Pi_{ij}^0 - \frac{3}{2} \right)\,,
\end{equation}
where $\vec{S}_i$ and $\vec{S}_j$ are SU(2) spin 1/2's, $\vec{T}_{ij}=\vec{S}_i+\vec{S}_j$ is the total
spin formed by the two spins $\vec{S}_i$ and $\vec{S}_j$, a (uniform) coupling constant is denoted
as $J$, and the  sum runs over all pairs of spins $i,j$ (or might be restricted to nearest neighbors on a 
given lattice).
Of course, the Heisenberg Hamiltonian is just a sum of projectors $\Pi_{ij}^0$ onto the pairwise
spin singlet state as can be easily seen
by rewriting the spin exchange $\vec{S}_i \cdot \vec{S}_j$ in terms of the total spin 
$\vec{T}_{ij}$. 
Antiferromagnetic coupling ($J>0$) favors an overall singlet state ($\vec{T}^2_{ij}=0$), 
while a ferromagnet coupling ($J<0$) favors the triplet state ($\vec{T}^2_{ij}=2$).

\begin{figure}[t]
  \center{
  \includegraphics[width=\columnwidth]{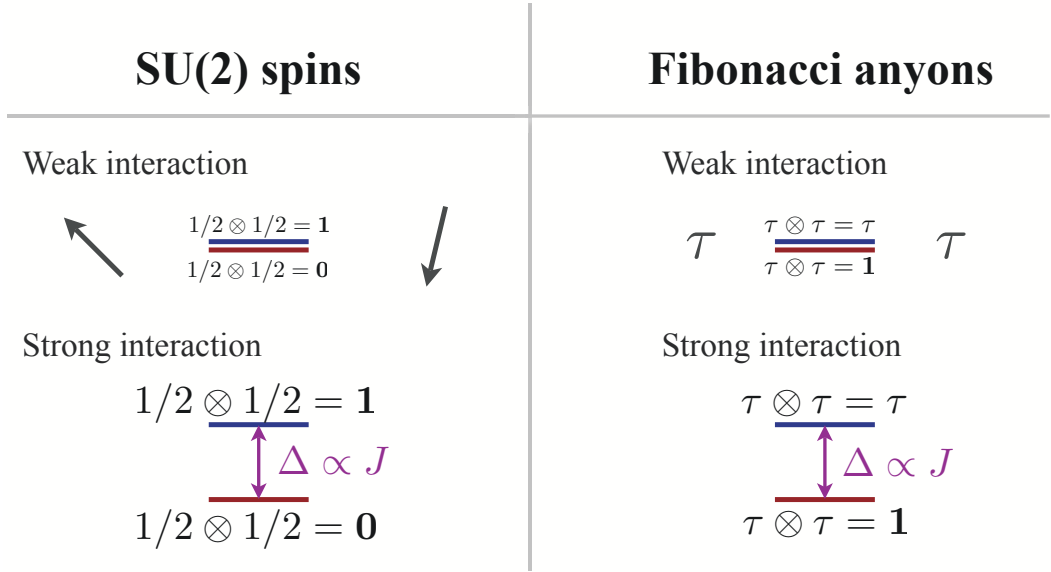}
  }
  \caption{
    (color online)
    The generalized Heisenberg model: 
    While for two weakly interacting SU(2) spin-1/2's (left panel) the total singlet and triplet 
    states are degenerate, strong interactions will lift this degeneracy. The Heisenberg
    Hamiltonian explicitly opens a gap of the order $\Delta \propto J$ between the two states.
    Applying a similar idea to the effect of interactions between two Fibonacci anyons
    (right panel), a generalized Heisenberg Hamiltonian will energetically distinguish the two 
    fusion outcomes.
   }
  \label{Fig:HeisenbergModel}
\end{figure}

In analogy, we can consider two Fibonacci anyons. If the two anyons are far apart and weakly
or non-interacting, then the two states that can be formed by fusing the two anyons will be 
degenerate. If, however, the anyons interact more strongly, then it is natural to assume that
the two fusion outcomes will no longer be degenerate and one of them is energetically favored.
We can thus generalize the Heisenberg Hamiltonian to anyonic degrees of freedom by expressing
it as a sum of projectors onto a given fusion outcome
\begin{equation}
\label{GoldenChainHamiltonian}
  \mathcal{H}^{\rm Fibonacci}_{\rm Heisenberg} = -J  \sum_{\langle ij \rangle} \Pi_{ij}^1\,
\end{equation}
where $\Pi_{ij}^1$ is a projector onto the trivial channel.

\subsection{The golden chain}
\begin{figure}[h]
  \center{
  \includegraphics[width=\columnwidth]{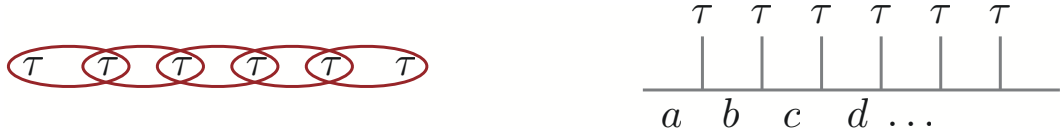}
  }
  \caption{(color online) The ``golden chain" of pairwise interacting Fibonacci anyons. 
                  The interactions are indicated by 
ellipses
around the anyons.
                  Our choice of fusion path basis is indicated in the right panel. }
\label{fig:chain}
\end{figure}

We will now explicitly derive the matrix representations for simple models of interacting
Fibonacci anyons. In the simplest model we consider a chain of Fibonacci anyons with 
nearest neighbor Heisenberg interactions as shown in Fig.~\ref{fig:chain}.  

\noindent
This Hamiltonian favors neighboring anyons to fuse into the trivial ({\bf 1}) channel by assigning an energy $-J$ to that fusion outcome. To derive the matrix representation in the fusion tree basis we first need to perform a basis change using the $F$-matrix
\begin{figure}[h]
  \center{
      \includegraphics[width=0.5\columnwidth]{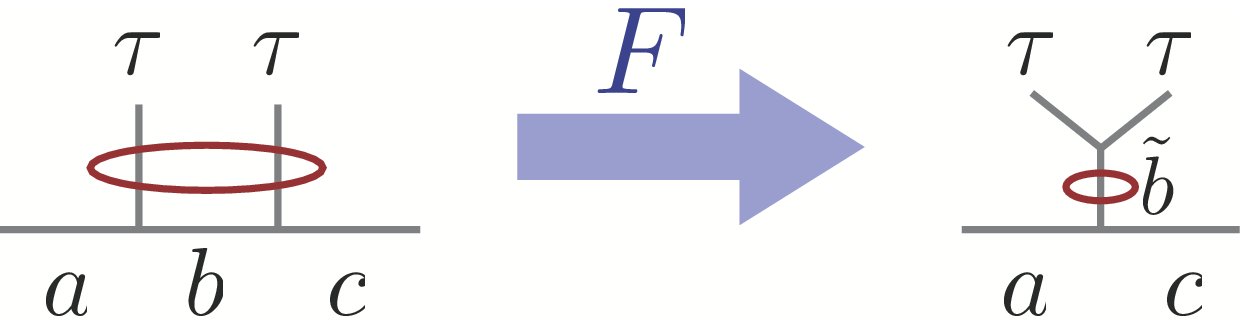}
  }
\end{figure}

\noindent
and then the Hamiltonian is just $F\Pi^1F$ with the projector $\Pi^1$ given by $\Pi^1={\rm diag}(1,0,0,1,0)$ in the basis (\ref{eq:basisafter}). Written explicitly this becomes
\begin{equation}
-J \left(F^{a\tau\tau}_c\right)^b_{1}\left(F^{a\tau\tau}_c\right)^1_{b'}.
\end{equation}
The matrix representation in the basis (\ref{eq:basisbefore}) then reads
 \begin{equation}
\label{SingletProjectorMatrixRepresentation}
-J\left(\begin{array}{ccccc}
1&&&& \\
&0&&& \\
&&0&&\\
&&&\ph^{-2}&\ph^{-3/2}\\
&&&\ph^{-3/2}&\ph^{-1}\end{array}
 \right) \,.
  \end{equation}

\clearpage
\subsection{Three-anyon fusion}
For the second model we include longer range interactions, preferring now to fuse {\em three} adjacent anyons into the trivial particle. For this we have to perform two basis changes to obtain the total charge of three anyons as shown here:

\begin{figure}[h]
  \center{
      \includegraphics[width=0.9\columnwidth]{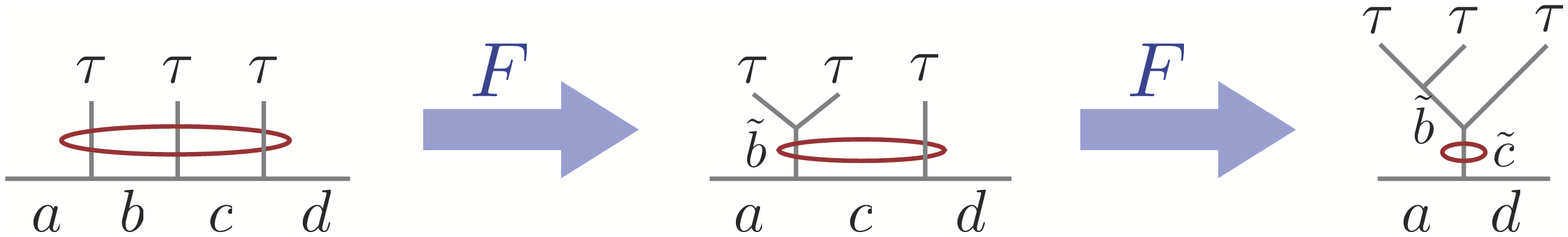}
  }
\end{figure}

The basis states of three anyons are given by the labelings of the four edges $|abcd\rangle$ between and adjacent to the three anyons:
  \begin{equation}
  \{|1\tau\tau 1\rangle, |1 \tau 1\tau\rangle,|1\tau\tau\tau\rangle,|\tau 1\tau 1\rangle,|\tau\tau\tau 1\rangle,|\tau\tau1\tau\rangle,\tau 1\tau\tau\rangle, |\tau\tau\tau\tau\rangle\} \,.
  \end{equation}
The first matrix is
  \begin{equation}
  F_1=\left(\begin{array}{cccccccc}
1&&&&&&& \\
&1&&&&&& \\
&&1&&&&&\\
&&&\ph^{-1}&\ph^{-1/2}&&&\\
&&&\ph^{-1/2}&-\ph^{-1}&&&\\
&&&&&1&& \\
&&&&&&\ph^{-1}&\ph^{-1/2}\\
&&&&&&\ph^{-1/2}&-\ph^{-1}
\end{array}\right)
  \end{equation}
  changing to the basis $ |a \tilde{b} c d\rangle$
  \begin{equation}
  \{|1\tau\tau 1\rangle, |1 1 1\tau\rangle,|1\tau\tau\tau\rangle,|\tau 1\tau 1\rangle,|\tau\tau\tau 1\rangle,|\tau \tau 1\tau\rangle,\tau 1\tau\tau\rangle, |\tau\tau\tau\tau\rangle\} \,,
  \end{equation}
and then a second basis change with
  \begin{equation}
  F_2=\left(\begin{array}{cccccccc}
1&&&&&&& \\
&1&&&&&& \\
&&1&&&&&\\
&&&1&&&&\\
&&&&1&&&\\
&&&&&\ph^{-1}&0&\ph^{-1/2}\\
&&&&&0&1&0 \\
&&&&&\ph^{-1/2}&0&-\ph^{-1}
\end{array}\right)
  \end{equation}
to a basis $ \ket{ a\tilde{b}\tilde{c}d }$
\begin{equation}
 \{|1\tau1 1\rangle, |1 1 \tau\tau\rangle,|1\tau\tau\tau\rangle,|\tau 1\tau 1\rangle,|\tau\tau\tau 1\rangle,|\tau 1\tau\tau\rangle,\tau\tau 1\tau\rangle, |\tau\tau\tau\tau\rangle\}.
\end{equation}
Combined with the projector $P_3={\rm diag}(1,0,0,0,0,1,0,0)$ the Hamiltonian matrix then becomes
  \begin{equation}
  H_3=-J_3F_1F_2P_3F_2F_1= -J_3\left(\begin{array}{cccccccc}
1&&&&&&& \\
&0&&&&&& \\
&&0&&&&&\\
&&&0&&&&\\
&&&&0&&&\\
&&&&&\ph^{-2}&\ph^{-2}&-\ph^{5/2}\\
&&&&&\ph^{-2}&\ph^{-2}&-\ph^{-5/2} \\
&&&&&-\ph^{-5/2}&-\ph^{-5/2}&-\ph^{-3}
\end{array}\right) \,.
\end{equation}
  
\subsection{Next-nearest neighbor interactions}
Another possibility for longer-range interactions is to fuse two particles at larger distance. Here we consider the case of next-nearest neighbor interactions. To determine the fusion result of these two anyons we first need to bring them to adjacent positions by braiding one of the particles with its neighbor as shown here:
\begin{figure}[h]
  \center{
      \includegraphics[width=0.7\columnwidth]{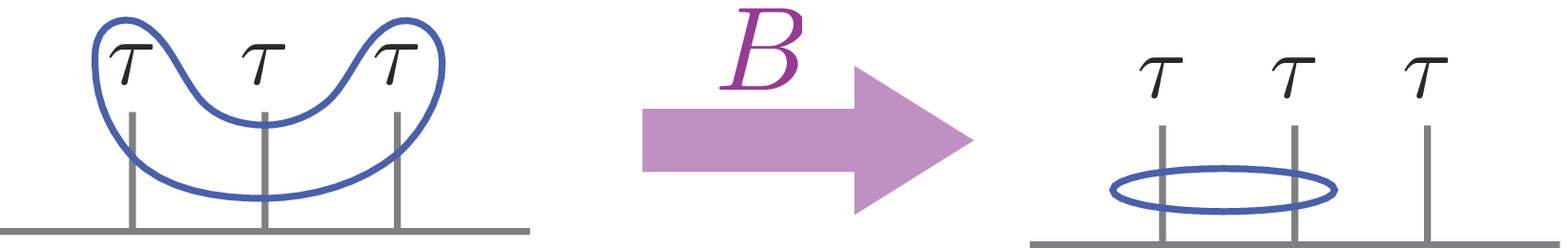}
  }
\end{figure}

\noindent
In contrast to Abelian particles, such as ordinary spins, for non-Abelian anyons it matters whether we braid the two anyons clockwise or counter-clockwise, since the braid matrix (for braiding clockwise) is different from its inverse (braiding anti-clockwise). Indicating the interaction of two anyons by loops around the two, the two ways of braiding correspond to the anyons fusing above or below the one between then as shown here:
\begin{figure}[h]
  \center{
      \includegraphics[width=0.25\columnwidth]{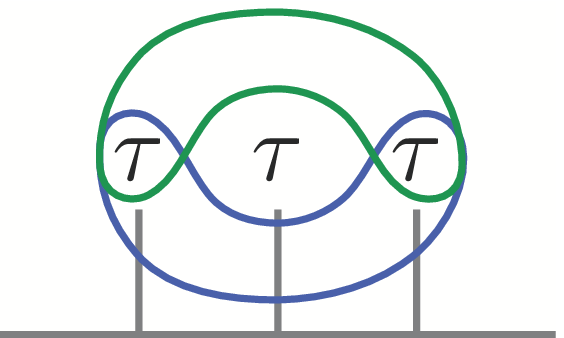}
  }
\end{figure}

\noindent
Using the same basis as above we obtain for the Hamiltonian for upper and lower interactions
\begin{eqnarray}
&&H_{2,{\rm above}} = BH_1B^{-1} = 
  H=FPF = \\ && -J_2\ph^{-3/2} \left(\begin{array}{cccccccc}
0&&&&&&& \\
&\ph^{-1/2}&-r^2&&&&&\\
&-r^{-2}&\ph^{1/2}&&&&&\\
&&&\ph^{-1/2}&-r^{2}&&&\\
&&&-r^{-2}&\ph^{1/2}&&&\\
&&&&&\ph^{-1/2}&-\ph^{-1/2}r^{4}&\ph^{-1}r^{2}\\
&&&&&-\ph^{-1/2}r^{-4}&\ph^{-1/2}&\ph^{-1}r^{-2}\\
&&&&&\ph^{-1}r^{-2}&\ph^{-1}r^{2}&\ph^{-3/2}
\end{array}
 \right)
 \nonumber
  \end{eqnarray}

and

  \begin{equation}
H_{2,{\rm below}} = B^{-1}H_1B =  H_{2,{\rm above}}^* \,,
  \end{equation}
where $B$ is the braid matrix acting on the first two anyone (first three labels) and $r=\exp(i\pi/5)$.
For the symmetrized model containing both terms we have
  \begin{eqnarray}
&&H_2=H_{2,{\rm below}}+H_{2,{\rm above}}=\\
&&-J_2\left(\begin{array}{cccccccc}
0&&&&&&& \\
&2\ph^{-2}&-\ph^{-5/2}&&&&&\\
&-\ph^{-5/2}&2\ph^{-1}&&&&&\\
&&&2\ph^{-2}&-\ph^{-5/2}&&&\\
&&&-\ph^{-5/2}&2\ph^{-1}&&&\\
&&&&&2\ph^{-2}&-\ph^{-1}&\ph^{-7/2}\\
&&&&&-\ph^{-1}&2\ph^{-2}&\ph^{-7/2}\\
&&&&&\ph^{-7/2}&\ph^{-7/2}&2\ph^{-3}
\end{array}
 \right)\nonumber \,.
\end{eqnarray}

As for SU(2) spin chains, the three-particle interaction can be written as a sum of nearest and 
next-nearest 
neighbor terms: For SU(2) spin-S this is
\begin{eqnarray}
H_3 &=& (\vec{S}_1+\vec{S}_2+\vec{S}_3)^2 = 3 S(S+1) {\bf 1} + 2 \vec{S}_1\vec{S}_2 + 2  \vec{S}_1\vec{S}_3 + 2 \vec{S}_2\vec{S}_3 \nonumber \\
&=&  3 S(S+1) {\bf 1}  + 2H_1^{12} + 2 H_1^{23} + 2 H_2^{13}
\end{eqnarray}
where the $H_1^{ij}$ indicates a nearest neighbor interaction between sites $i$ and $j$ and $H_2^{ij}$ a next-nearest neighbor one. Mapping an $H_3$ chain to a chain containing both $H_1$ and $H_2$ terms we find $J_1=4J_3$ and $J_2=2J_3$. The $H_3$ chain thus corresponds to the Majumdar-Ghosh chain \cite{MajumdarGhosh} with an exact ground state of singlet dimers.

Similarly we find for Fibonacci anyons
\begin{eqnarray}
H_3 &=& - {\bf 1}  -\ph^{-2} H_1^{12} -\ph^{-2} H_1^{23} -\ph^{-1}  H_2^{13}
\end{eqnarray}
and thus $J_1=-2\ph^{-2}J_3$ and $J_2=-\ph^{-1}J_3$. Again the pure $H_3$ chain corresponds to the Majumdar-Ghosh chain with an exact ground state of anyonic dimers fusing into the trivial particle
\cite{CollectiveStates}.
 
Generalizations of the next nearest neighbor interaction to longer distances is straightforward, the only complication arises from the various possible orientations of the braids, giving $2^{r-1}$ different terms at distance $r$.

\subsection{The two-leg ladder}

\begin{figure}[h]
  \center{
  \includegraphics[width=\columnwidth]{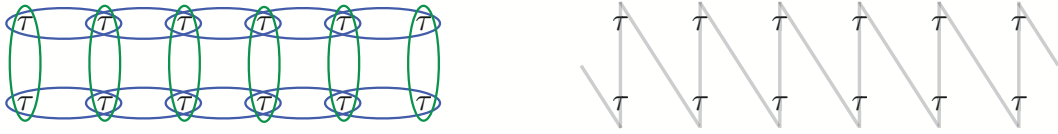}
  }
  \caption{(color online) A two-leg ladder model, consisting of two chains of Fibonacci anyons. 
                  Pairwise interactions are indicated by ellipsis around the anyons.
                  Our choice of fusion path is indicated by a line in right panel. }
\label{fig:ladder}
\end{figure}

The final model we will present is a two-leg ladder consisting of two coupled chains. Unlike the case of the chain, where it was natural to just use the standard fusion tree as basis, there are several natural choices here. Choosing the zig-zag fusion path indicated by a line in Fig.~\ref{fig:ladder} minimizes the interaction range on the fusion tree. 

Distinguishing rung interactions with coupling constant $J_\perp$ and chain interactions with coupling $J$, we see that the rung terms are just nearest neighbor interactions and the chain terms are next nearest neighbor interactions above or below the other anyon for the upper and lower chain respectively:
\begin{equation}
H_{\rm ladder} = J_\perp \sum_i H_1^{2i-1,2i} + J\sum_i H_{2,{\rm below}}^{2i-1,2i+1} + J \sum_i H_{2, {\rm above}}^{2i,2i+2} \,.
\end{equation}

Having established the explicit matrix representations of these Hamiltonians we can now analyze
their ground states thereby addressing the original question of what kind of collective ground states a set of Fibonacci anyons will form in the presence of these interactions.



\section{Alternative SU(2)$_k$ formulation}

Before turning to the collective ground states of the Hamiltonians introduced above,
we will describe an alternative reformulation of the
 ``golden chain",
Eq. (\ref{GoldenChainHamiltonian}),
which allows
 us to generalize 
the latter and, in particular, allows us to view the  golden chain and its variants as certain 
deformations of ordinary SU(2) spin chains.
Technically, this  generalization is based on connections with the famous work by V. F. R. 
Jones~\cite{RefTemperleyLiebForSuTwoLevelK} on representations of the Temperley-Lieb algebra
\cite{TemperleyLieb}.

Specifically, this reformulation is based on the fusion rules of
the anyon theory known by the name
\lq SU(2) at
level $k$\rq,
which in symbols we denote as SU(2)$_k$, for $k=3$.  
For arbitrary integer values of the parameter $k$ there exist particles labeled by
\lq angular momenta\rq
\ 
$j=0, 1/2, 1, 3/2, ..., k/2$.
The result of fusing two particles with \lq angular momenta\rq
\ $j_1$ and $j_2$ yields particles 
 with \lq angular momenta\rq \ $j$ 
where
$j =$
$ |j_1-j_2|, |j_1-j_2|+1, ..., \min\{j_1+j_2, k-(j_1+j_2)\}$,
each occuring with multiplicity $N_{{j_1} {j_2}}^j=1$.
When $k \to \infty$
this 
represents 
ordinary SU(2)  spins,
for which all possible values of (ordinary) SU(2) angular
momenta $j=0, 1/2, 1, 3/2, ....$ appear,
and the fusion rule turns into the
 ordinary
angular momentum coupling
rule.
Finite values of $k$
represent a  \lq quantization\rq \ of SU(2),
amounting to the indicated  truncation of the range
of \lq angular momenta\rq.
Thus, our reformulation will allow us to view the ``golden chain"
as  a deformation of the ordinary  SU(2) Heisenberg spin chain
by a parameter $1/k$.
For the special case where $1/k = 1/3$
we obtain, as we will now briefly review, 
the chain of Fibonacci anyons introduced
in Eq. (\ref{GoldenChainHamiltonian}) above.

Our reformulation of the ``golden chain" is based
on two simple observations:
(i) first, one recalls that at $k=3$  (where the four particles $j=0,1/2,1,3/2$ exist)
the fusion rule of the particle with $j=1$ is
$ 1 \otimes 1 = 0 \oplus 1$ (all entries label \lq angular momenta\rq),
which is identical to that of the Fibonacci anyon $\tau$, e.g. 
$ \tau \otimes \tau = {\bf 1} \oplus \tau$.
Note that the trivial particle,
previously denoted by ${\bf 1}$,  is now denoted
by \lq angular momentum\rq \ $j=0$.
(ii) Secondly, we recall that the particle $j=3/2$ 
can be used to map the particles one-to-one into
each other (it represents what is known as
an automorphism of the fusion algebra),
i.e.
\begin{equation}
\label{FusionWithThreeHalf}
3/2 \otimes 0 = 3/2,
\quad
 3/2 \otimes 1/2 = 1,
\quad
 3/2 \otimes 1 = 1/2,
\quad
 3/2 \otimes 3/2 = 0.
\end{equation}
We denote this automorphism by
\begin{equation}
\label{DEFhat}
3/2 \otimes j = \hat j: \qquad j \to \hat j \,.
\end{equation}
Note that when $j$ is an integer, then $\hat j$ is a half-integer,
and vice-versa.

Using the observation
 (i) above we may first write an arbitrary state vector in the 
Hilbert space which can be represented pictorially as 
\begin{figure}[h]
  \center{
      \includegraphics[width=0.5\columnwidth]{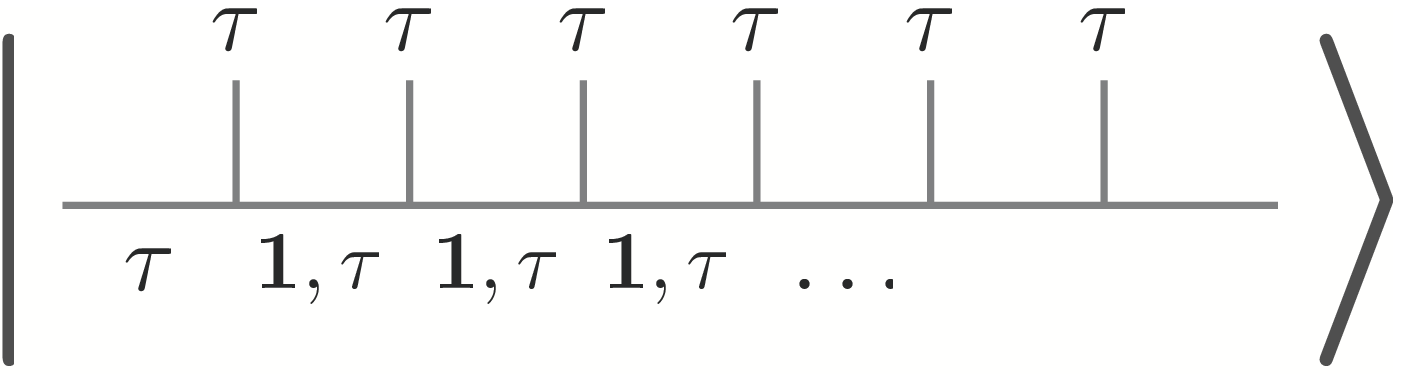} \,,
  }
\end{figure}

\noindent
in the new notation using the 'angular momenta' as follows
\begin{figure}[h]
  \center{
      \includegraphics[width=0.5\columnwidth]{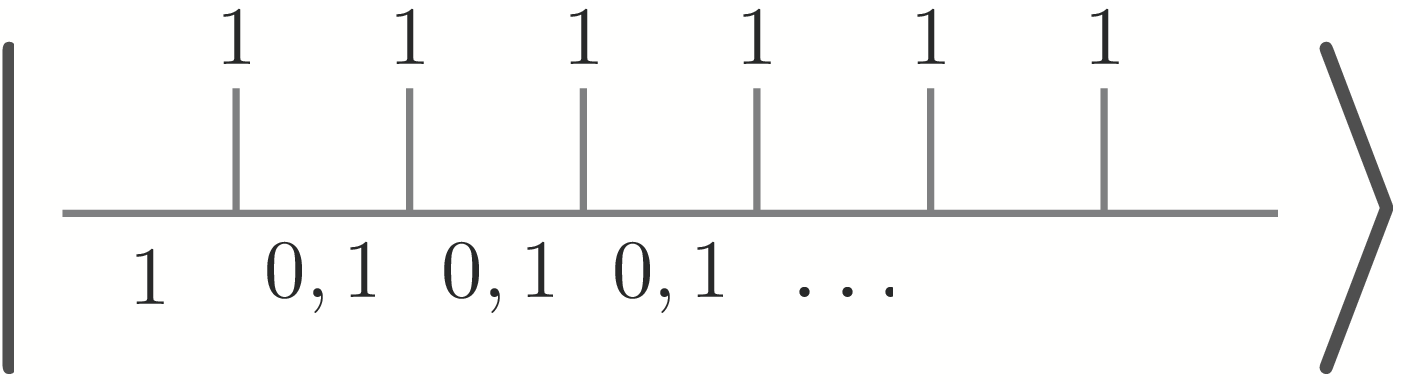} \,.
  }
\end{figure}

\begin{figure}[b]
  \center{
      \includegraphics[width=\columnwidth]{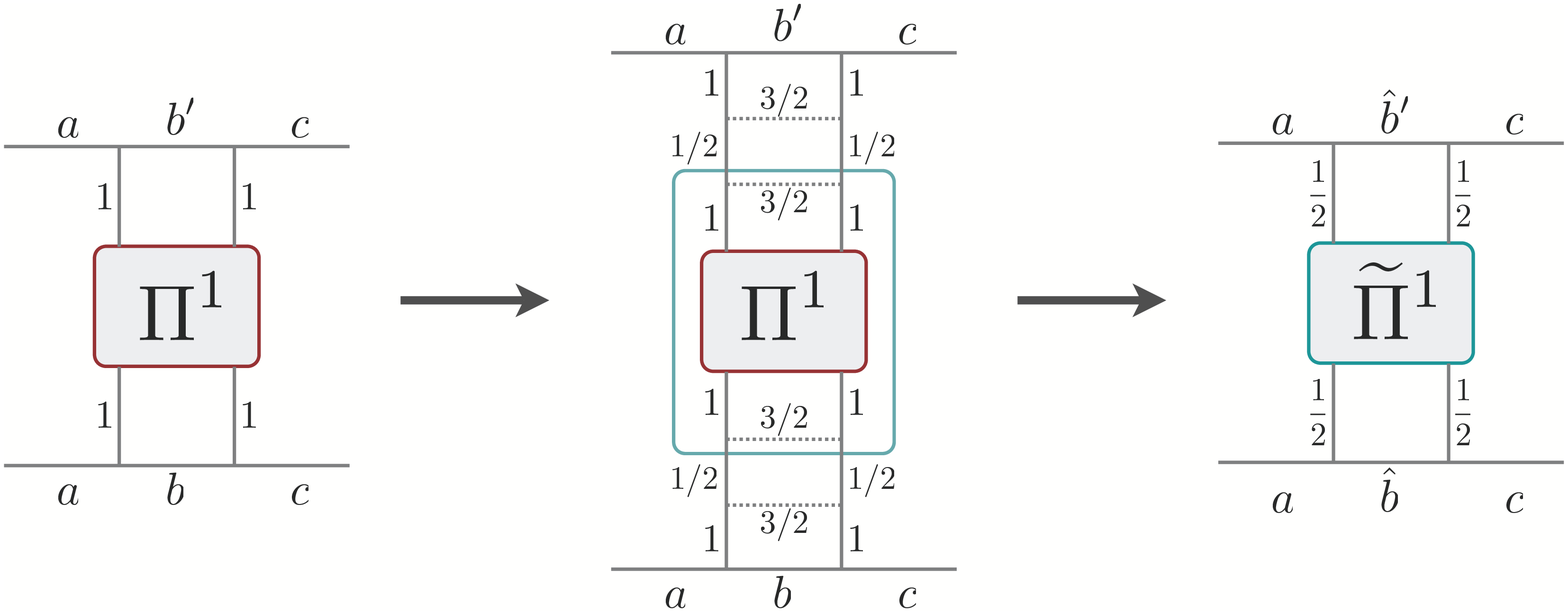}
  }
  \caption{(color online) Transformation of the projector $\Pi^1$.}
  \label{Fig:ProjectorEquation}
\end{figure}

Let us now consider the matrix element of the projector 
[discussed in Eq.s (\ref{GoldenChainHamiltonian})
-(\ref{SingletProjectorMatrixRepresentation})
above]
\begin{equation}
\label{MatrixElementProjector}
\langle ... a', b', c' ...| \Pi^1 | ... a, b, c ... \rangle
\end{equation}
where $ ... a', b', c', ... a, b, c, ... \in \{ j=0, j=1 \}$.
A moment's reflection shows that in fact $a'=a$ and $c'=c$.
This matrix element is graphically depicted in the leftmost picture in 
Fig.~\ref{Fig:ProjectorEquation}, where the direction from the right (ket) 
to the left (bra) in (\ref{MatrixElementProjector})
is now drawn vertically upwards from bottom to top in the Figure.

We may now perform a number of elementary steps using
a $1\times 1$-dimensional $F$-matrix (which is just a number) 
involving the Abelian particle $j=3/2$.
Specifically, we perform the steps depicted as
\begin{figure}[h]
  \center{
      \includegraphics[width=\columnwidth]{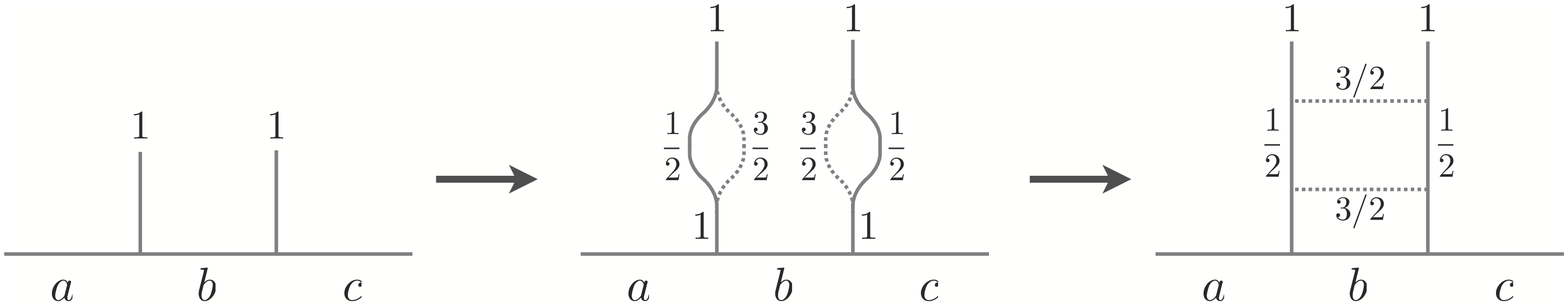}
  }
\end{figure}

\noindent
on both,  the upper and the lower legs of the projector in 
Fig.~\ref{Fig:ProjectorEquation}, where in the last step we use
\begin{figure}[h]
  \center{
      \includegraphics[width=\columnwidth]{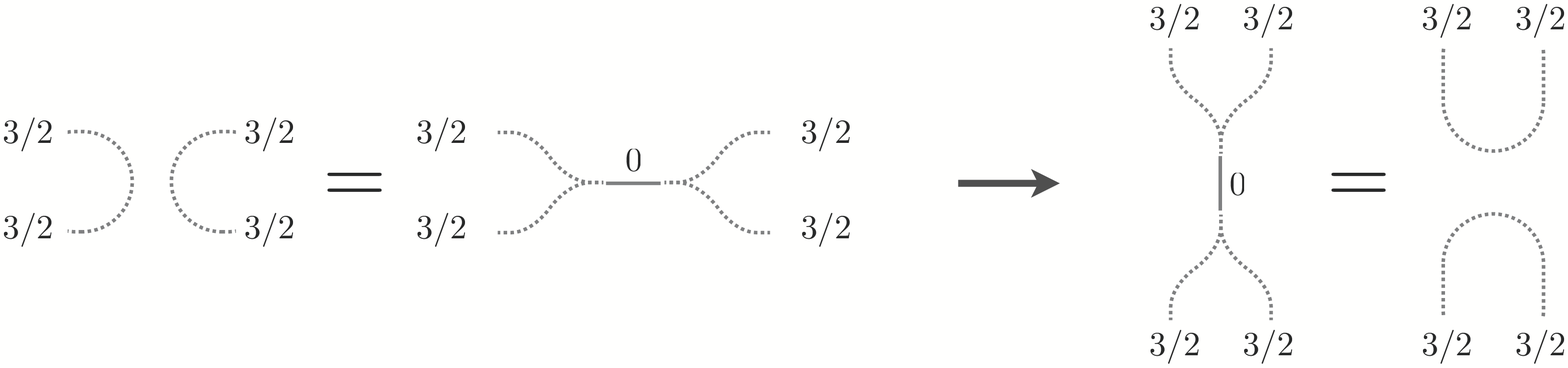}
  }
\end{figure}

\noindent
Subsequently we use,
in the diagram in the middle of the projector equation Fig.~\ref{Fig:ProjectorEquation}, 
the transformation
\begin{figure}[h]
  \center{
      \includegraphics[width=\columnwidth]{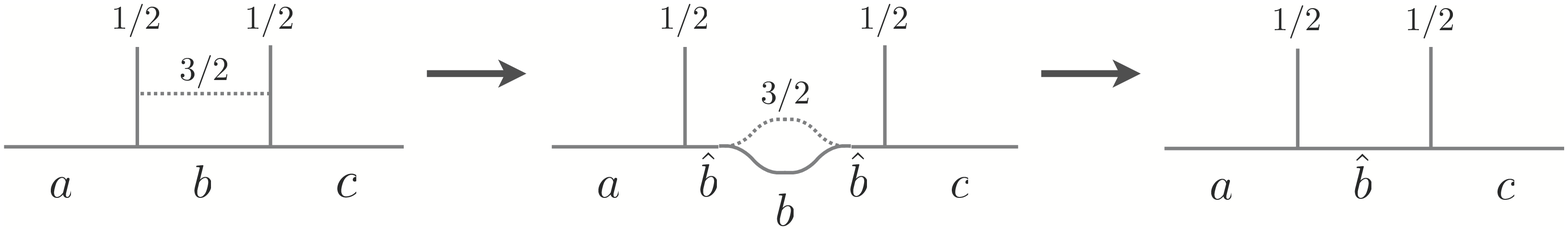}
  }
\end{figure}

\noindent
where in the first step we have applied the relation
\begin{figure}[h]
  \center{
      \includegraphics[width=0.55\columnwidth]{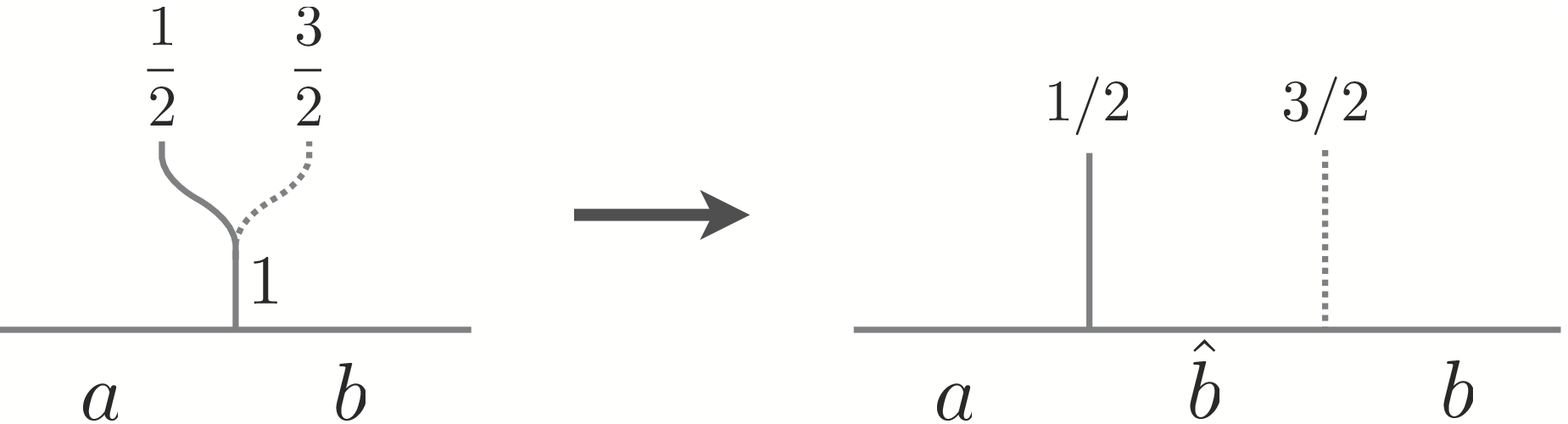}
  }
\end{figure}

\noindent
The projector $\tilde{\Pi^1}$ with four $j=1/2$ particle legs
(depicted in the rightmost diagram of Fig.~\ref{Fig:ProjectorEquation}),
which we obtain at the end of this sequence of elementary steps is the central object
of interest to us.

As it is evident from the rightmost picture in Fig.~\ref{Fig:ProjectorEquation},
the projector $\tilde{\Pi^1}$ acts on basis states
of the form
\begin{equation}
\label{HilbertSpaceQuantumSpinoneHalf}
 \ket{... a, {\hat{b}}, c, {\hat{d}}, e, ...} \,,
\end{equation}
that is on states in which integer and half-integer labels 
are strictly alternating.
Such basis states can then be depicted as
\begin{figure}[h]
  \center{
      \includegraphics[width=0.5\columnwidth]{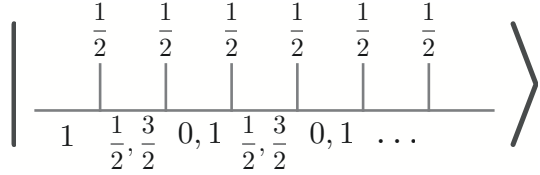} \,.
  }
  \caption{`Angular momentum' description of the wavefunction for the Fibonacci theory SU(2)$_3$.}
  \label{Fig:SU(2)3-WaveFunction}
\end{figure}

The projector $\tilde{\Pi^1}$ that we have thereby obtained is
a central and well known object of study in Mathematics: 
Specifically,
let us form the operator
\begin{equation}
\label{JonesTemperleyLiebGenerator}
e_i := d \  \tilde{\Pi^1}_i
\end{equation}
where $d = \ph = (\sqrt{5}+1)/2$ is the golden ratio.
The index $i$ indicates the label along the
fusion chain basis of \lq angular momenta\rq \ in
(\ref{HilbertSpaceQuantumSpinoneHalf})
on which it acts. For example, the operator $\Pi^1$
in (\ref{MatrixElementProjector})
as drawn in Fig.~\ref{Fig:ProjectorEquation} acts on the label $b$
which is, say, in the $i^{th}$ position in the chain of symbols
characterizing the state.
With these notations, the operators $e_i$
are known to provide a representation
of the Temperley-Lieb algebra \cite{TemperleyLieb}

$$
e_i^2 = d \, e_i \,,
$$
$$
e_i e_{i\pm1} e_i = e_i \,,
$$
$$
[ e_i, e_j ] = 0, \quad |i-j| \geq 2 \,.
$$
with $d$-isotopy parameter $d = \ph = (\sqrt{5}+1)/2$.
The so-obtained representation of the Temperley-Lieb algebra
is only one in an infinite sequence with different values of
the $d$-isotopy parameter, 
discovered by V. F. R.  Jones\cite{RefTemperleyLiebForSuTwoLevelK}.
Specifically, in our
notations, this infinite sequence 
is labeled by the integer $k$,
specifying the $d$-isotopy parameter
$d=2 \cos[\pi/(k+2)]$.
For any integer value of
$k$, the Jones representation is given by
the matrix elements of the operator $e_i$ 
in the basis of type (\ref{HilbertSpaceQuantumSpinoneHalf}).
Recall from the discussion at the beginning of this paragraph
that for general values of the integer $k$
the labels $ ... a, {\hat{b}}, c ...$
are the \lq angular momenta\rq \ in
the set
$j = 0, 1/2, 1, 3/2, ..., k/2$. Thus it is in general
more convenient to write these basis states as
\begin{equation}
\label{HilbertSpaceQuantumSpinoneHalfGeneralk}
\ket{ \ldots j_{i-1}, j_i , j_{i+1} \ldots}
\end{equation}
with $j_{i-1}, j_i , j_{i+1} \in \{  0, 1/2, 1, 3/2, ..., k/2\}$;
the sequence must satisfy
the condition that
$j_i$ is contained in the fusion product of $j_{i-1}$ with an \lq angular momentum\rq \ $1/2$,
$j_{i+1}$ is contained in the fusion product of $j_{i}$ with \lq angular momentum\rq \ $1/2$,
etc. -- this generalizes  the states depicted in Fig.~\ref{Fig:SU(2)3-WaveFunction} 
(which is written for $k=3$).
Within this notation the matrix elements of the operator $e_i$ are
(compare \ref{MatrixElementProjector})
\begin{equation}
\label{MatrixElementProjectorGeneralk}
\langle ... j_{i-1}', j_i' , j_{i+1}' ...  | e_i |
 ... j_{i-1}, j_i , j_{i+1} ...  \rangle
=
\delta_{j_{i-1}', j_{i-1}}
\ 
\delta_{j_{i+1}', j_{i+1}}
\
\delta_{j_{i-1}, j_{i+1}}
\
\sqrt{
{
S^0_{j_i} S^0_{j_i'}
\over
S^0_{j_{i-1}} S^0_{j_{i+1}}
} \,,
}
\end{equation}
where
\begin{equation}
\label{ModularSMatrix}
S^{j'}_j
=
\sqrt{ 2 \over (k+2)}
\sin[
\pi
{ (2j+1) (2j'+1)
\over
k+2}
] \,.
\end{equation}

This, then, defines a generalization of the original ``golden chain",
for which
$k=3$,
to arbitrary values of $k$. 
It is interesting to note
that in the limit $k\to \infty$, the so-defined generalized anyonic chain
turns precisely into the ordinary SU(2) spin-1/2 Heisenberg chain.
Therefore, these generalized ``golden chains"
(for arbirary integer values of $k$)
provide, as alluded to in the introduction of this section,
a certain generalization of the ordinary SU(2) spin-1/2 Heisenberg chain.
Therefore, it is natural to ask questions about the behavior
of ordinary spin-1/2 chains, in the context of these generalized
golden chains. Indeed, as we have described in various publications
\cite{GoldenChain,CollectiveStates,SU2k}
and as reviewed in this note, many of the physics questions
familiar from the ordinary Heisenberg chain find their
mirror-image in the behavior of our  generalized golden chains.
The following section is intended to give a brief flavor of
these results and parallels.

Before we proceed, let us briefly mention that for \lq antiferromagnetic\rq \ 
coupling of the generalized spins of  the golden chain (for general $k$),
meaning that projection onto the singlet state 
(trivial anyon
fusion channel)
between two spins
is energetically favored, 
one obtains a gapless theory which turns out to be precisely
the $(k-1)$-th minimal model of conformal field theory\cite{BPZ}
of central charge
$c=1-6/[(k+1)(k+2)]$.
 (For
$k=2$ this is the Ising model, for $k=3$ the tricritical Ising model,
and so on.) In the opposite,  \lq ferromagnetic\rq \  case, 
meaning the case
where the projection
onto the \lq generalized triplet\rq \ state of two neighboring
generalized spins is energetically favored, one obtains another
well known sequence of gapless models: these are the $Z_k$
parafermionic conformal field theories\cite{ZamolodchikovFateevParafermions},
of central charge $c=2(k-1)/(k+2)$, 
which have more recently attracted attention
in quantum Hall physics as potential candidates
for certain  non-Abelian quantum Hall states, known
as the Read-Rezayi states \cite{ReadRezayi}. 
For a summary see Table I.

\begin{table}
\begin{center}
\begin{tabular}{c||c|c}
level $k$ & AFM & FM \\
\hline
\multirow{2}{*}{2} & $c=1/2$ & $c=1/2$ \\
   & Ising & Ising \\
\hline
\multirow{2}{*}{3}  & $c=7/10$ & $c=4/5$ \\
 & tricritical Ising & 3-state Potts \\
\hline
\multirow{2}{*}{4}  & $c=4/5$ & $c=1$ 
\\
 & tetracritical Ising & 
$Z_4$-parafermions
\\
\hline
\multirow{2}{*}{5}  & $c=6/7$ & $c=8/7$ 
\\
 & pentacritical Ising &
$Z_5$-parafermions 
 \\
\hline
\multirow{2}{*}{$k$}  & $c=1-6/[(k+1)(k+2)]$ & $c=2(k-1)/(k+2)$ \\
 & $k^{th}$-multicritical Ising & $Z_k$-parafermions \\
\hline
\multirow{2}{*}{$\infty$} & $c=1$ & $c=2$ \\
 & Heisenberg AFM & Heisenberg FM \\
\end{tabular}
\end{center}
\caption{The
 gapless (conformal field) theories describing the generalized 
spin-1/2 chain
of anyons with
SU(2)$_k$  non-Abelian statistics.
}
\label{tab:models}
\end{table}


\section{Collective ground states}

In this final section we will round off the manuscript by shortly reviewing some recent numerical and
analytical work analyzing the collective ground states formed by a set of Fibonacci anyons in the presence of the generalized Heisenberg interactions introduced in the previous section.

For the ``golden chain" investigated in Ref.~\citen{GoldenChain} , a one-dimensional arrangement of Fibonacci anyons with nearest neighbor interaction terms, it has been shown
(as already mentioned above)
that the system is 
gapless
-- independent of which fusion channel, the trivial channel ($\bf 1$) for
antiferromagnetic coupling, or 
the $\tau$-channel for ferromagnetic coupling,
is energetically favored by the pairwise fusion. 
The finite-size gap $\Delta(L)$ for a system with $L$ Fibonacci anyons vanishes as 
$\Delta(L) \propto (1/L)^{z=1}$ with dynamical critical
exponent $z=1$, indicative of a conformally invariant energy spectrum. 
The two-dimensional conformal field  theories describing the system have central charge $c=7/10$ for 
antiferromagnetic interactions and $c=4/5$ for ferromagnetic interactions, 
respectively,
corresponding to the entry for $k=3$ in Table I.
In fact, a direct connection to the corresponding two-dimensional classical models, the tricritical Ising
model for $c=7/10$ and the three-state Potts model for $c=4/5$, has been made:  Realizing that
(as reviewed briefly in the previous section)
the non-commuting local operators of the ``golden chain" Hamiltonian form 
a well known representation\cite{RefTemperleyLiebForSuTwoLevelK}
of the Temperley-Lieb algebra\cite{TemperleyLieb} 
(with $d$-isotopy parameter $d=\ph$), it has been shown\cite{GoldenChain}
that the Hamiltonian of this quantum chain corresponds precisely to 
(a strongly anisotropic version of) the transfer matrix of the
integrable restricted-solid-on-solid (RSOS) lattice 
model\cite{ABF}, thereby mapping the 
anyonic quantum chain
exactly onto the tricritical Ising and three-state Potts critical points of the generalized 
hard hexagon model \cite{GoldenChain,Huse82}.  
A corresponding exact  relationship holds in fact true for
the chains at any value of the integer $k$ (Table I).

\begin{figure}[t]
 \center{
  \includegraphics[width=0.8\columnwidth]{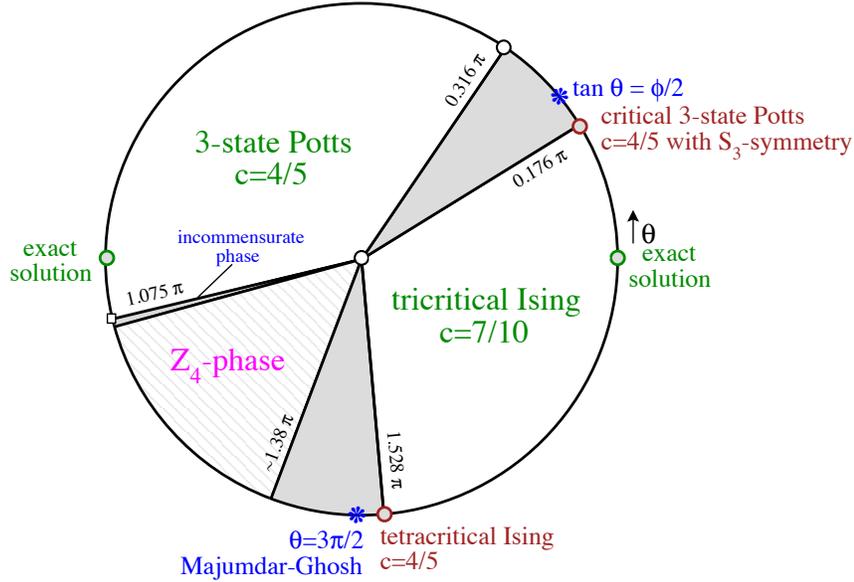}}
  \caption{
    (color online)
    The phase diagram of the anyonic Majumdar-Ghosh chain, 
    for Fibonacci anyons,
    as presented in 
    Ref.~\citen{CollectiveStates}.
    The exchange couplings are parametrized on the circle by an angle $\theta$,      
    with a pairwise fusion term $J_2=\cos\theta$
    and a three-particle fusion term $J_3=\sin\theta$.
    Besides extended critical phases around the exactly solvable points ($\theta=0,\pi$) that can
    be mapped to the tricritical Ising model and the 3-state Potts model, there are two gapped
    phases (grey filled).
    The phase transitions (red circles) out of the tricritical Ising phase
    exhibit higher symmetries and are both described by CFTs with central charge $4/5$.
    In the gapped phases exact ground states are known at the positions marked by the stars.
    In the lower left quadrant a small sliver of an incommensurate phase occurs and a phase
    which has $Z_4$-symmetry.  These latter two phases also appear to be
    critical.
   }
  \label{Fig:J3-Chain:PhaseDiagram}
\end{figure}

While this correspondence of the ``golden chain" 
with
the special critical points of the classical models
might
initially
 seem accidental,
it turns out that the quantum system exhibits
in fact
an additional topological symmetry
that actually stabilizes the gaplessness of the quantum system, and protects it 
from local perturbations
which would generate a gap.
In particular, it was shown that all translational-invariant relevant operators that appear in the quantum
system, e.g. the two thermal operators
of the tricritical Ising model (antiferromagnetic case)
$\epsilon, \epsilon'$ with scaling 
dimensions $1/5$ and $6/5$ respectively, are 
forbidden by this topological symmetry\cite{GoldenChain}.

A more detailed connection to the underlying two-dimensional classical models has been made
by considering the effect of a competing next-nearest neighbor interaction or, 
equivalently, a three-anyon fusion term in the anyonic analog of the 
Majumdar-Ghosh chain \cite{CollectiveStates},
derived explicitly above. 
The rich phase diagram of this model is reproduced in Fig.~\ref{Fig:J3-Chain:PhaseDiagram}.
Besides extended critical phases around the ``golden chain" limits ($\theta=0,\pi$) for which an exact
solution is known, there are two gapped phases (grey filled) with distinct ground-state degeneracies and
well-defined quasiparticle excitations in the spectrum. 
The tricritical Ising phase ends in higher-symmetry critical endpoints, with an $S_3$-symmetry at
$\theta=0.176\pi$ and 
an Ising tetracritical point
at $\theta=-0.472\pi$. The transitions at these points 
spontaneously break the topological symmetry and in the case of the tetracritical point also the
translational symmetry thereby giving rise to two-fold and four-fold degenerate ground states in the
adjacent gapped phases, respectively. For an in-depth discussion of this phase diagram 
we refer to Ref.~\citen{CollectiveStates}.

Finally, the effect of random interactions on chains of (Fibonacci) anyons has been 
studied in Refs.~\citen{RandomChain,RandomChain2}. 
For random, `antiferromagnetic' interactions the random system is found to flow to strong disorder 
and the infinite randomness fixed point is described by a generalized random singlet phase
\cite{RandomChain}.
For a finite density of `ferromagnetic' interactions an additional `mixed phase' infinite randomness
fixed point is found \cite{RandomChain2}.

\section*{Acknowledgments}
We thank E. Ardonne, A. Feiguin, M. Freedman, C. Gils, D. Huse, and A. Kitaev for many 
illuminating discussions and joint work on a number of related publications.
We further acknowledge stimulating discussions with P. Bonderson, N. Bonesteel, L. Fidkowski, 
P. Fendley, C. Nayak, G. Refael, S.H. Simon, J. Slingerland, and K. Yang.
 

\end{document}